\documentclass[prx,floatfix,superscriptaddress,longbibliography,notitlepage, twocolumn, 10.5pt]{revtex4-2}
\usepackage[utf8]{inputenc}
\usepackage{graphicx}
\usepackage[dvipsnames]{xcolor}
\usepackage{float}
\usepackage[citecolor=RawSienna, colorlinks=true, linkcolor=MidnightBlue, urlcolor=MidnightBlue, linktocpage=true]{hyperref}
\usepackage[T1]{fontenc}
\usepackage{bbm}
\usepackage{amsmath}
\usepackage{lmodern}
\usepackage{amsfonts}
\usepackage{physics}
\usepackage{mathtools}
\usepackage{tikz}
\usepackage{mathtools}
\usepackage{appendix}
\usetikzlibrary{arrows.meta}
\usepackage{dsfont}

\definecolor{colblue}{RGB}{31, 119, 180}    
\definecolor{colred}{RGB}{214, 39, 40}      

\tikzstyle{vertex_mt}=[circle, draw=colblue, very thick, text=colblue, font=\normalsize, minimum size=40pt, inner sep=0pt]
\tikzstyle{vertex_mt2}=[circle, fill=white, draw=colblue, very thick, text=colblue, font=\normalsize, minimum size=40pt, inner sep=0pt]
\tikzstyle{edge} = [draw,ultra thick,-]

\begin{document}


\title{Enhanced analogue quantum computation with a logically-encoded XX gadget}

\author{Robert J.~Banks}
\affiliation{London Centre for Nanotechnology, University College London, Gower Street, WC1E 6BT, London, United Kingdom}
\author{Natasha Feinstein}
\affiliation{London Centre for Nanotechnology, University College London, Gower Street, WC1E 6BT, London, United Kingdom}
\author{Roopayan Ghosh}
\affiliation{School of Basic Sciences, Indian Institute of Technology, Bhubaneswar, 752050, India.}
\affiliation{Department of Physics and Astronomy, University College London, Gower Street, WC1E 6BT, London, United Kingdom}
\author{Sebastian Schulz}
\affiliation{J\"ulich Supercomputing Centre, Institute for Advanced Simulation,\\Forschungszentrum J\"ulich, 52425 J\"ulich, Germany}

\author{Kristel Michielsen}
\affiliation{J\"ulich Supercomputing Centre, Institute for Advanced Simulation,\\Forschungszentrum J\"ulich, 52425 J\"ulich, Germany}
\affiliation{Universit\"at zu K\"oln, 50923 K\"oln, Germany}
\author{Sougato Bose}
\affiliation{Department of Physics and Astronomy, University College London, Gower Street, WC1E 6BT, London, United Kingdom}
\author{P.~A.~Warburton}
\email{p.warburton@ucl.ac.uk}
\affiliation{London Centre for Nanotechnology, University College London, Gower Street, WC1E 6BT, London, United Kingdom}
\date{\today}

\date{\today}


\begin{abstract}




The introduction of two-body $XX$ interactions has been shown to improve the performance of quantum annealing in solving certain combinatorial optimisation problems and would enable the simulation of a broader class of many-body Hamiltonians. Due to experimental constraints, such interactions are typically not natively available on analogue quantum devices. To overcome this limitation we introduce the truss gadget. This is an exact construction that generates two-body $XX$ interactions using only one-body $X$ and $Z$ fields and two-body $ZZ$ interactions which are readily accessible on, for instance, superconducting quantum annealers. We experimentally validate the construction by reproducing the dynamics of a target $XX$ Hamiltonian on a trapped-ion quantum processor. We then apply the gadget to an illustrative set of instances of the weighted maximum-independent-set problem on a superconducting quantum annealer. We empirically determine an approximate thousand-fold improvement in the probability of finding a state close to the ground state for these instances, over implementation without an $XX$ interaction.
\end{abstract}

\maketitle

\section{ Introduction \label{sec:introduction} }

The classical Ising model is known to provide one of the simplest and most fundamental frameworks for understanding phase transitions and collective behaviour in interacting many-body systems. From a computational perspective, the associated decision problem of determining whether an Ising Hamiltonian admits a configuration below a given energy threshold is NP-complete, while finding its ground state is NP-hard~\cite{Istrail_00, Luc14}. This computational complexity makes the Ising Hamiltonian a natural choice for encoding a broad class of combinatorial optimization problems. In particular, canonical NP-hard problems such as the travelling salesperson, maximum satisfiability, and minimum graph-colouring problems can be mapped onto Ising Hamiltonians with only polynomial overhead. Quantum annealing exploits controlled quantum fluctuations to navigate the energy landscape of such Ising-encoded optimisation problems. Experimental quantum annealing has been pursued on several physical platforms, most notably superconducting qubits \cite{Johnson2011} 
and neutral atoms \cite{Glaetzle_2017,Eba22,de_Oliveira_2025}. For superconducting qubits the problem Hamiltonian and accompanying driver is closely described by the transverse-field Ising model (TFIM) within the qubit subspace.

In the TFIM quantum fluctuations are introduced by a transverse field that does not commute with the longitudinal fields and Ising interactions of the classical model. In quantum annealing the transverse field is monotonically reduced to zero while the longitudinal terms in the Hamiltonian (which encode the problem of interest) increase from zero. The transverse field opens a non-zero spectral gap, $\Delta$, between the ground state and the first excited state within the same symmetry sector. In the closed system setting, the adiabatic principle therefore guarantees the possibility of remaining in the ground state given that the fields and interactions are varied sufficiently slowly~\cite{Far00}. The adiabatic condition imposes a runtime that depends on the minimum spectral gap as well as on matrix elements involving derivatives of the time-dependent Hamiltonian, with the precise scaling depending on the annealing schedule \cite{Jansen2007, Roland2002}. In many cases, this gap minimum closes exponentially with system size. This can happen as a result of a first order phase transition between the paramagnetic phase, associated with a dominant transverse field, and some localised phase associated with the longitudinal terms \cite{seki2012,Alb19,Nishimori_2017}. These exponentially closing gaps can also arise from transitions between different localised states that are separated by a large Hamming distance \cite{Ami09,Alt10}. When such exponentially small gaps occur along the annealing path, the corresponding adiabatic runtime becomes exponential in system size, posing a fundamental challenge to the efficiency of quantum annealing for these instances.

Including higher-order transverse interactions, $X_1\,X_2\,\dots\,X_k$, introduces correlated spin-flips, which can enhance the tunnelling rate between the states on either side of a gap minimum by comparison with including only the standard one-local transverse-field. In particular, two-body $XX$ interactions are well studied in the literature and have been found to result in gap enhancement in a variety of settings. For the annealing of the analytically tractable p-spin and Hopfield models, such interactions have been shown to remove a first-order phase transition \cite{seki2012,Alb19,Nishimori_2017}. Numerical studies have also demonstrated that such couplings have the capacity for gap enhancement, and thus more effective annealing, in various other settings \cite{Alb19,Hormozi_2017,Mehta_2021,Takada_2020,Fei24,gho24,nut24}. Some studies have drawn attention to the advantages offered by $XX$ interactions introduced with the \textit{opposite} sign to the local $X$ fields, resulting in a total Hamiltonian which is non-stoquastic \cite{seki2012,Nishimori_2017,choi2022}. This class of Hamiltonian is harder to simulate classically, with, for example, quantum Monte Carlo simulations encountering the sign problem \cite{gupta2019}. On the other hand, stoquastic $XX$ interactions have generally been found, both numerically and analytically, to outperform their non-stoquastic counterparts with respect to gap enhancement for certain classes of problems~\cite{Hormozi_2017,Mehta_2021,nut24,Crosson2020}. The current state of both superconducting and neutral-atom implementations of quantum annealing does not however allow the direct implementation of $k$-body transverse interactions (with $k>1$) simultaneously with two-qubit longitudinal interactions. This raises a natural question: can the spectral benefits of higher-order transverse interactions be obtained using only the native interaction resources of existing quantum-annealing platforms?

Motivated by the interaction set available on current quantum-annealing hardware, several gadget constructions have been proposed for the implementation of two-body \(XX\) interactions using only one-local \(X\)-fields together with longitudinal fields and interactions. These constructions involve different trade-offs in the strength and accuracy of the resulting effective interaction and in their hardware requirements. Perturbative gadgets operate by implementing a penalty that energetically separates a sub-space which encodes the logical basis states in such a way that the desired effective logical interactions arise from the implemented Hamiltonian~\cite{Kempe2006, Jor08}. Two such gadgets for implementing an $XX$ coupling were introduced in \cite{banks2025gadgetssimulatingnonnativexx}, including one which can be implemented using just the TFIM interactions. A downside of such gadgets is the trade-off between the accuracy of the effective \(XX\) interaction and its strength: increasing the energy penalty improves the separation between the logical and excited subspaces, but suppresses the resulting effective interaction strength. There has also been a proposal for a \textit{non}-perturbative approach to implementing $k-$local transverse interactions through chain gadgets constructed from interlocking $3-$ or $5-$body $Z$ interactions \cite{Headley_2026}. Here the effective interactions are not suppressed by an energy penalty. However, the higher-order $Z$ terms, as well as an additional requirement for complex logical operations limits the implementability of these gadgets on the devices of interest. More recently, a method by which inhomogeneous driving could be used to emulate a fully connected $XX$ graph has been proposed \cite{dadgar2026emulatingxxcatalystsquantum}. This approach becomes exact in the large system limit. 

Here we introduce the \textit{truss gadget} \cite{trusspatent} which has the following benefits over the aforementioned approaches. Unlike perturbative gadgets, its effective \(XX\) interaction is exact by construction and does not rely on an asymptotically large energy penalty. Since it does not use a penalty term to energetically separate a logical sub-space, there is no sub-space to leak out of, even once open system effects are introduced. Finally, the gadget itself can be implemented using only the TFIM with homogeneous driving. The qubit overhead of the gadget is at worst quadratic in system size. If $3-$ and $4-$ local $Z$ interactions are available, then the hardware overhead is exactly one additional qubit per chain of $XX$ interactions. We experimentally demonstrate the truss construction on a trapped-ion quantum processor by reproducing the dynamics of a target logical \(XX\) Hamiltonian. We then demonstrate its utility as an annealing catalyst by applying it to a weighted maximum-independent-set problem on a superconducting quantum annealer, where it enhances the probability of obtaining the ground state.

The layout of this paper is as follows. We introduce the truss gadget and the intuition behind it in section \ref{sec:gadget}. In section \ref{sec:experiment} we give experimental results on quantum hardware which confirm that the gadget does indeed implement an $XX$ interaction (section \ref{sec:IonQ}) and can be used to enhance the probability of finding the ground state in an example quantum annealing problem (section \ref{sec:D-Wave}). Section \ref{sec:conclusion} presents our conclusions.

\section{ The truss gadget \label{sec:gadget} }

\begin{figure}
    \centering
    
    \begin{tikzpicture}[scale=0.8, auto,swap]
    \node[font=\Large] at (-4, 12) {\textbf{(a)}};
    \node[font=\Large] at (2, 12) {\textbf{(b)}};
        \foreach \pos/\name in { {(-2.4-0.25,2.4-1.2)/1}, {(0-0.25,4.8-1.2)/2}, {(-2.4-0.25,7.2-1.2)/3}, {(0-0.25,9.6-1.2)/4}, {(-2.4-0.25,12-1.2)/5}}
                \node[vertex_mt] (\name) at \pos {$-d_\name X_\name$};
        \foreach \source/ \dest  in {1/2, 2/3, 3/4, 4/5}
                \path[edge] (\source) -- node {$\text{\textit{\normalsize{h}}}_\text{\normalsize{\source}}$}(\dest);
        
        \path[edge] (1) -- node[left] {$\text{\textit{\normalsize{J}}}_\text{\normalsize{1,2}}$}(3);
        \path[edge] (2) -- node {$\text{\textit{\normalsize{J}}}_\text{\normalsize{2,3}}$}(4);
        \path[edge] (3) -- node[left] {$\text{\textit{\normalsize{J}}}_\text{\normalsize{3,4}}$}(5);
        
        \draw[colred, ultra thick] (4,0) .. node[left] {$\text{\textit{\normalsize{-d}}}_\text{\normalsize{2}}$} controls (3,2) .. (4,4);
        \draw[colred, ultra thick] (4,4) .. node[left] {$\text{\textit{\normalsize{-d}}}_\text{\normalsize{3}}$} controls (3,6) .. (4,8);
        \draw[colred, ultra thick] (4,8) .. node[left] {$\text{\textit{\normalsize{-d}}}_\text{\normalsize{4}}$} controls (3,10) .. (4,12);
        
        \draw[ultra thick] (4,0) .. node[right] {$\text{\textit{\normalsize{J}}}_\text{\normalsize{1,2}}$} controls (5,2) .. (4,4);
        \draw[ultra thick] (4,4) .. node[right] {$\text{\textit{\normalsize{J}}}_\text{\normalsize{2,3}}$} controls (5,6) .. (4,8);
        \draw[ultra thick] (4,8) .. node[right] {$\text{\textit{\normalsize{J}}}_\text{\normalsize{3,4}}$} controls (5,10) .. (4,12);
        
        \node[vertex_mt2, align=center] (6) at (4,0){$h_1 Z_1$\\[2pt]$-d_1 X_1$};
        \node[vertex_mt2] (7) at (4,4) {$h_2 Z_2$};
        \node[vertex_mt2] (8) at (4,8) {$h_3 Z_3$};
        \node[vertex_mt2, align=center] (9) at (4,12) {$h_4 Z_4$\\[2pt]$-d_5 X_4$};
    \end{tikzpicture}
    
    \caption{The truss gadget described in section \ref{sec:gadget} for $n=5$ physical qubits. Qubits are represented by blue nodes. (a) shows the physical qubits and (b) shows the equivalent logical spin-chain. Black edges denote $ZZ$ interactions. Red edges denote $XX$ interactions. Each edge is labelled with the strength of the interaction. The text inside each node shows the local field per qubit.}
    \label{fig:truss}
\end{figure}
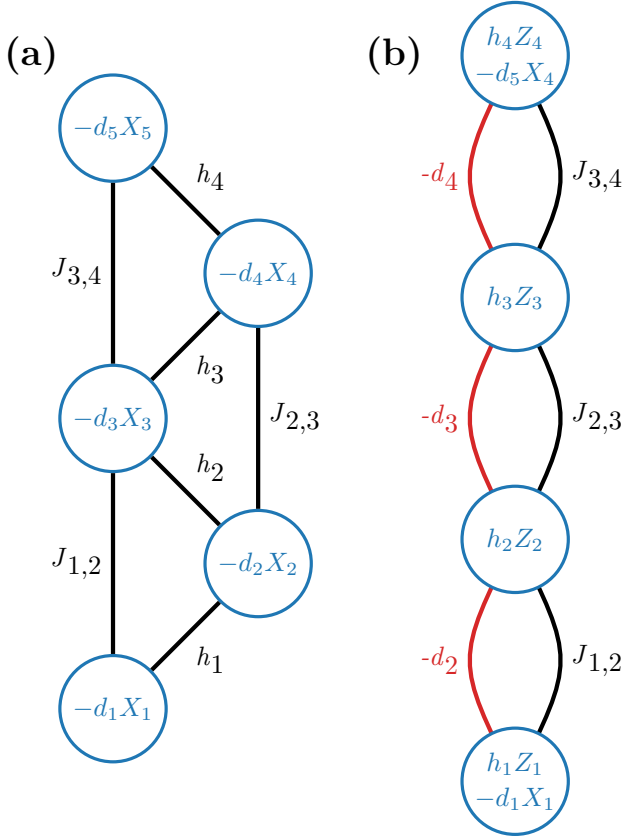

The truss gadget works by encoding logical qubits into pairs of physical qubits. The basic intuition behind the gadget is that by flipping a \textit{single physical spin} associated with two logical qubits we change the parity of two spin pairs. As a result, \textit{two logical qubits} are flipped, emulating an $XX$ interaction.  

First we consider the gadget between a single pair of logical qubits. The two logical qubits are encoded into three physical qubits. The physical Hamiltonian to be realised for a single effective $XX$ interaction is:
\begin{equation}
    \hat{H}_\text{Phys}^\text{Drive}=-\sum_{i=1}^3 d_i \hat{X}_i 
\end{equation}
Each logical qubit is encoded as the parity of two of the physical qubits:
\begin{align}
    \tilde{Z}_1&\equiv \hat{Z}_1 \hat{Z}_2 \label{eq:dimz1}\\
    \tilde{Z}_2&\equiv \hat{Z}_2 \hat{Z}_3.\label{eq:dimz2}
\end{align}
The resulting logical Hamiltonian under this encoding is:
\begin{equation}
    \label{eq:log_truss}
    \tilde{H}_\text{Logical}^\text{Drive}=-d_1\tilde{X}_1-d_3\tilde{X}_2-d_2\tilde{X}_1\tilde{X}_2,
\end{equation}
realising an effective $\pm XX$ interaction. Each term in the effective Hamiltonian shown in Eq.~\ref{eq:log_truss} can be independently tuned through inhomogeneous driving. The implementation of the logical single-body Pauli-Z's are shown in Eq.~\ref{eq:dimz1} and Eq.~\ref{eq:dimz2}. The logical two-body $ZZ$ interaction is given by:
\begin{align}
    \tilde{Z}_1\tilde{Z}_2&\equiv \hat{Z}_1 \hat{Z}_2 \hat{Z}_2 \hat{Z}_3 \\
    &\equiv \hat{Z}_1 \hat{Z}_3\label{eq:dimz12}.
\end{align}
Hence, the gadget is able to realise local $X$ and $Z$ fields on a pair of logical qubits as well as both $XX$ and $ZZ$ interactions between them. 

Next we consider how to implement effective $XX$ interactions between multiple pairs of logical qubits. Multiple gadgets can be linked together in a one-dimensional structure, at the expense of no longer realising single-body $X$ driving on the internal qubits. The resulting arrangement of physical qubits resembles a Warren truss \cite{Connor_2016}, hence the name of the gadget. The truss gadget is illustrated in Fig.~\ref{fig:truss}. Considering first only the physical transverse fields, the Hamiltonian with $n$ qubits is given by
\begin{equation}
    \hat{H}_\text{Phys}^\text{Drive}=-\sum_{i=1}^{n} d_i \hat{X}_i
\end{equation}
Each logical qubit is encoded as the parity of two adjacent physical qubits, \textit{i.e.}
\begin{equation}
    \tilde{Z}_i\equiv \hat{Z}_{i} \hat{Z}_{i+1}.
    \label{eq:phystologic}
\end{equation}
The corresponding logical Hamiltonian with $n-1$ qubits is given by:
\begin{equation}
    \tilde{H}_\text{Logical}^\text{Drive}=-d_1\tilde{X}_1-d_{n}\tilde{X}_{n-1}
    -\sum_{i=1}^{n-2} d_{i+1} \tilde{X}_i\tilde{X}_{i+1}.
\end{equation}
Longitudinal interactions can now be added to the logical Hamiltonian by noting that products of adjacent logical Pauli-Z's can be implemented with two-body physical operators:
\begin{equation}
    \prod_{r=p}^{p+q}\tilde{Z}_r\equiv \hat{Z}_{p} \hat{Z}_{p+q+1}.
    \label{eq:cong_z}
\end{equation}

To summarize, the $n$-qubit physical Hamiltonian
\begin{multline}
    \hat{H}_\text{Phys}^\text{Truss}=-\sum_{i=1}^{n} d_i \hat{X}_i 
    +\sum_{i=1}^{n-1} h_{i} \hat{Z}_{i} \hat{Z}_{i+1}\\
    +\sum_{i=1}^{n-2} J_{i,i+1} \hat{Z}_{i} \hat{Z}_{i+2} ,
    \label{eq:H_truss_phy}
\end{multline}
realises the chain of $n-1$ logical qubits:
\begin{multline}
    \tilde{H}_\text{Logical}^\text{Truss}=-\sum_{i=1}^{n-2} d_{i+1} \tilde{X}_i\tilde{X}_{i+1}-d_1 \tilde{X}_1-d_{n} \tilde{X}_{n-1}\\
    +\sum_{i=1}^{n-1} h_i \tilde{Z}_i+\sum_{i=1}^{n-2} J_{i,i+1} \tilde{Z}_i\tilde{Z}_{i+1},
    \label{eq:H_truss_log}
\end{multline}
where each qubit is encoded as the parity of two physical qubits.

The realisation of a logical $ZZ$ interaction between two non-adjacent qubits in the truss gadget requires a physical four-body $ZZZZ$ interaction. Further to this, single-qubit logical $X$ terms can be encoded as
\begin{equation}
    \tilde{X}_i \equiv \prod_{j=1}^i\hat{X}_j,
\end{equation}
up to the global spin-flip symmetry in the gadget. To couple from the truss to unencoded qubits requires a physical three-body $ZZZ$ interaction. Despite this, as we will show in section \ref{sec:D-Wave}, by keeping only the set of physically realisable interactions on current hardware (\textit{i.e.} those in Eq.~\ref{eq:H_truss_phy}), the truss gadget can lead to enhanced computational performance for combinatorial optimisation problems that are otherwise hard for quantum annealing. 

\section{Experimental benchmarking of the truss gadget}
\label{sec:experiment}

In this section we report two different experimental benchmarks of the truss gadget. Our goals are firstly to establish that the gadget results in dynamics that are consistent with an $XX$ interaction and secondly to demonstrate that the effective $XX$ interactions can provide the same advantages as a native $XX$ coupling in a combinatorial optimisation context. For the latter goal we make use of the superconducting quantum annealer manufactured by D-Wave \cite{Johnson2011} since it is specifically designed for combinatorial optimisation. For the former goal, however, we would require mid-anneal measurements (\textit{i.e.} measurements of the state of the annealed system at times shorter than the annealing time) to establish the full dynamical time-dependence of the quantum state. Mid-anneal measurements \cite{deshpande2026analogdigitalquantumcomputingquantum} are non-trivial on the D-Wave annealer since the rate at which the system can be quenched is in general lower than the rate at which the system evolves under the action of the transverse field. To study the dynamical behaviour we therefore implement a Trotterized version of the truss gadget on a gate-based quantum computer (specifically the ion-trap quantum computer manufactured by IonQ~\cite{Chen2024}) and measure its evolution. 

We begin, in section \ref{sec:IonQ}, by using the truss gadget to implement a closed ring of qubits on the IonQ Forte gate-based ion-trap quantum computer, comparing the dynamics of the Trotterized evolution to a closed-system simulation of a ring of genuine $XX$ interactions. Next, in section \ref{sec:D-Wave}, we consider a quantum annealing setting in which an $XX$ coupling is expected to result in an enhancement to the final ground-state fidelity. By implementing the truss on the D-Wave Advantage system, we confirm that the effective $XX$ interaction is indeed able to enhance the success of the algorithm.
\begin{figure*}
    \centering
    \includegraphics[width=0.98\linewidth]{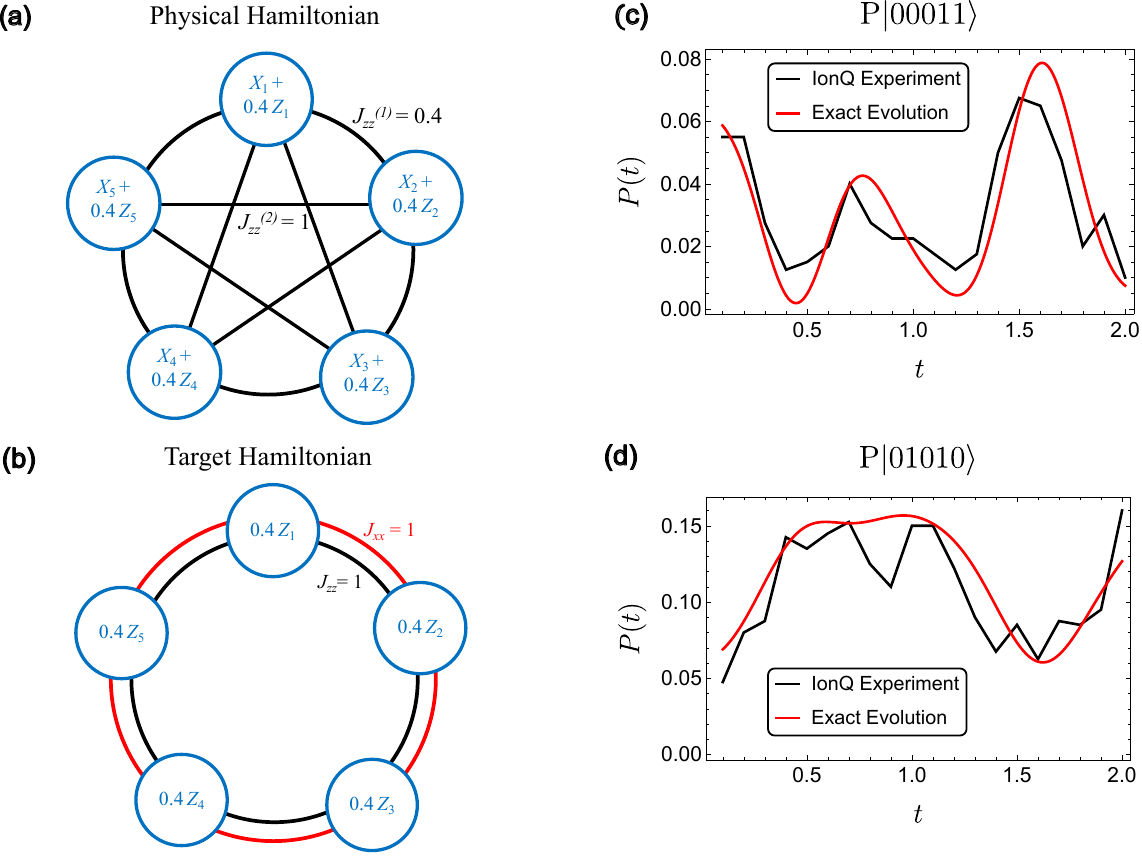}    
   \caption{(a) Physical and (b) target Hamiltonians used in the ion trap experiment. Qubits are in blue while $ZZ$ and $XX$ interactions are black and red, respectively. The local fields per qubit are shown in blue. In the physical system all nearest-neighbour interactions $J_{zz}^{(1)}$ are of the same magnitude, as are all next-nearest-neighbour interactions $J_{zz}^{(2)}$. In the target system all $ZZ$ and $XX$ interactions are of the same magnitude. Time evolution of the occupation probabilities for the computational basis states $|00011\rangle$ (c) and $|01010\rangle$ (d). The exact continuous-time evolution (red) is compared with the corresponding measurements obtained from the truss implementation on the IonQ Forte processor (black). The close agreement demonstrates that the truss gadget faithfully reproduces the target target dynamics on quantum hardware.}
    \label{fig:singleprob}
\end{figure*}
\subsection{Dynamics of the truss gadget on an ion-trap quantum computer}
\label{sec:IonQ}
We verify the experimental realization of the truss construction on an
IonQ Forte trapped-ion processor by comparing its dynamics with exact
continuous-time evolution and a direct implementation of the target $XX$
Hamiltonian. Real-time and Trotterized Hamiltonian dynamics have previously
been demonstrated on trapped-ion processors, including IonQ hardware
\cite{Mueller2023,Ltstedt2024,Sarkar2026}; here we specifically test whether
the truss construction reproduces the target $XX$ dynamics on
hardware.

We consider a closed loop of $n=5$ physical qubits, extending the truss gadget in Eq.~\ref{eq:H_truss_phy} to include all nearest and next-nearest $\hat{Z}\hat{Z}$ on a closed-loop. Using the truss gadget this can be mapped to dynamics restricted to a fixed parity sector,
containing $2^{n-1}$ states and hence $n-1=4$ independent logical degrees of
freedom. The full details on the mapping can be found in Appendix~\ref{app:loop_details}. We choose the physical Hamiltonian with nearest-neighbour Ising
interaction $J_{zz}^{(1)}=0.4$, next-nearest-neighbour interaction
$J_{zz}^{(2)}=1.0$, and transverse-field strength $d=1.0$. Under the parity
transformation discussed in the previous section, these parameters correspond
to a target Hamiltonian with longitudinal field $h=0.4$, nearest-neighbour
Ising interaction $J_{zz}=1.0$, and unit-strength nearest-neighbour $XX$
interaction. The physical and corresponding target Hamiltonians are shown
schematically in Figs.~\ref{fig:singleprob}(a) and
\ref{fig:singleprob}(b), respectively.

For the truss implementation, the evolution takes place directly in the
physical Hilbert space, with the variables reconstructed after
measurement through the parity mapping
$\tilde{Z}_i=\hat{Z}_i\hat{Z}_{i+1}$. In contrast, the direct $XX$
implementation represents the dynamics within a restricted subspace
of the physical Hilbert space. The two implementations therefore realize the
same logical Hamiltonian through distinct physical constructions,
providing a direct comparison of the truss and physical $XX$ realizations.

The initial states are prepared accordingly. For the physical implementation,
the quench starts from the paramagnetic ground state of the transverse-field
Hamiltonian,
\begin{equation}
|\psi_{\mathrm{phys}}(0)\rangle
=
|+\rangle_{\hat{Z}}^{\otimes n},
\qquad
|+\rangle_{\hat{Z}}
=
\frac{|0\rangle_{\hat{Z}}+|1\rangle_{\hat{Z}}}{\sqrt{2}},
\end{equation}
prepared experimentally by applying Hadamard gates to all qubits initialized in $|0\rangle$. For the direct $XX$ implementation, the
corresponding initial state is the equal superposition over the allowed
even-parity sector,
\begin{equation}
|\psi_{\mathrm{enc}}(0)\rangle
=
\frac{1}{2^{(n-1)/2}}
\sum_{\substack{x_1,\ldots,x_n\\
x_1\oplus\cdots\oplus x_n=0}}
|x_1\cdots x_n\rangle_{\tilde{Z}}.
\end{equation}
This state contains $2^{n-1}$ configurations and represents the paramagnetic state within the parity-constrained Hilbert space. This parity-constrained initial state is prepared by applying Hadamard gates to four physical qubits, followed by four CNOT gates that compute their parity onto the fifth qubit.

We benchmark four descriptions of the resulting dynamics: (i) exact
continuous-time evolution, (ii) a noiseless digital simulation of the target
Hamiltonian using nearest-neighbour $XX$ interactions, (iii) an experimental
realization of the physical Hamiltonian using the truss gadget, and (iv) a
direct experimental realization of the target Hamiltonian using
nearest-neighbour $XX$ interactions. The exact evolution is obtained from
\begin{equation}
|\psi(T)\rangle=e^{-iHT}|\psi(0)\rangle,
\end{equation}
and serves as the reference for the digital and experimental
implementations.

The digital evolution is performed using an alternating first-order
Suzuki--Trotter decomposition that becomes equivalent to a symmetric
second-order (Strang) decomposition over pairs of successive time steps.
The Hamiltonian is separated into two mutually non-commuting sectors,
\begin{equation}
H=H_{X/XX}+H_Z,
\end{equation}
where $H_{X/XX}$ denotes the transverse $\hat{X}$ contribution in the
physical implementation and the nearest-neighbour
$\tilde{X}\tilde{X}$ interaction in the logical implementation. The
diagonal sector $H_Z$ contains the Ising interactions and longitudinal
field. Since all terms within each sector mutually commute, each sector
can be implemented exactly within a single Trotter layer.

\begin{figure*}
    \centering
    \includegraphics[width=0.98\linewidth]{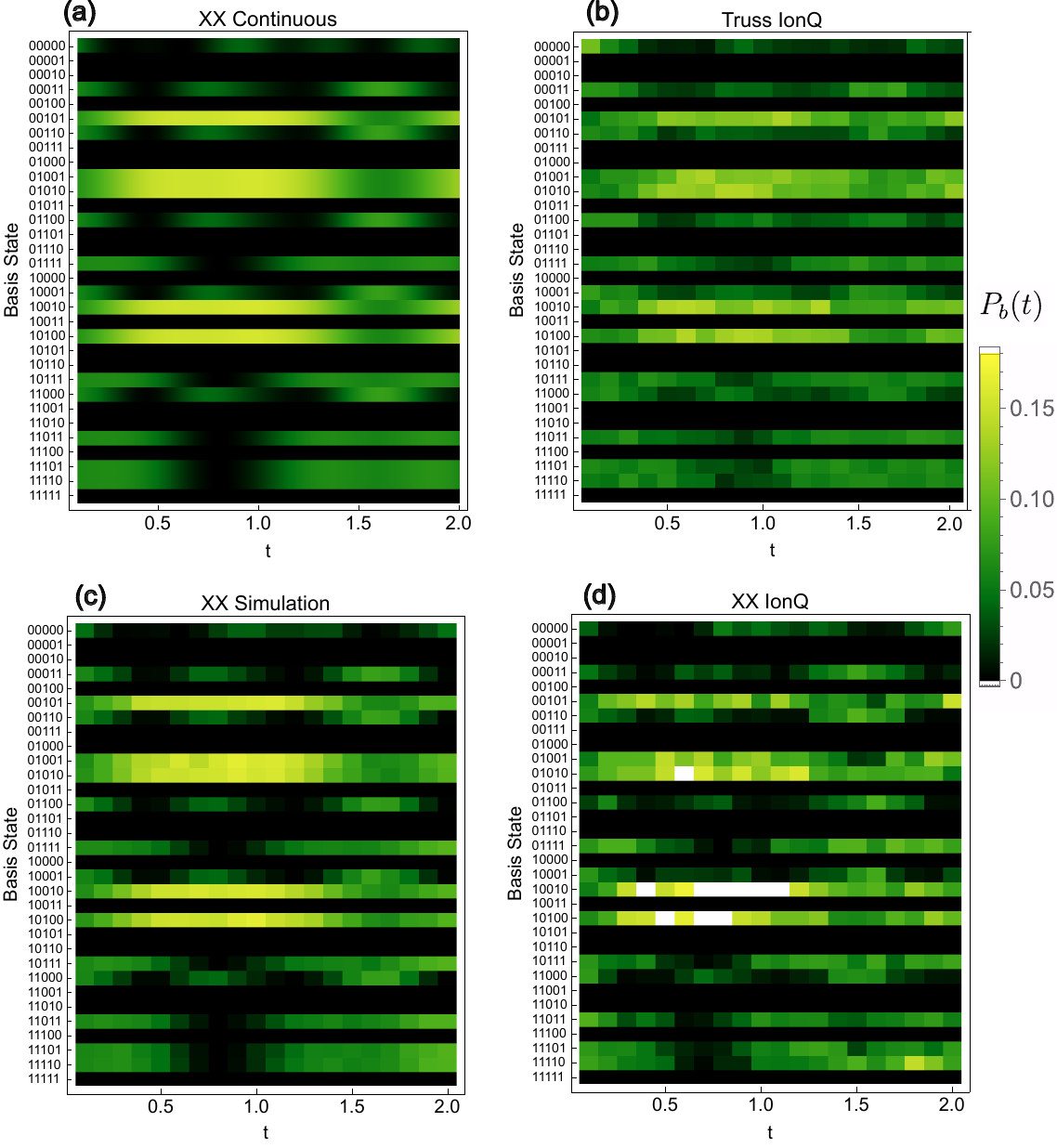}
\caption{Time evolution of the probability distribution over the computational basis states for a five-qubit system. Each row corresponds to a computational basis state (the down  and up spin states are denoted by \(0\) and \(1\), respectively), while the horizontal axis denotes the evolution time. The colour represents the probability $P_b(t)=|\langle b|\psi(t)\rangle|^2$ of measuring the computational basis state $b$, with darker colours indicating lower probabilities and brighter yellow indicating higher probabilities, as shown by the accompanying colour bar. Panel (a) shows the exact continuous-time evolution obtained by numerically solving the Schr\"odinger equation. Panel (b) shows the corresponding implementation of the physical Hamiltonian using the truss gadget on the IonQ Forte processor, with the measured bit strings mapped back to the same basis as the target Hamiltonian through the parity transformation. Panel (c) shows the corresponding noiseless digital simulation of the target Hamiltonian employing nearest-neighbour $XX$ interactions, while panel (d) shows its implementation directly on the IonQ Forte processor. The comparison
illustrates the agreement between the exact, digital, and experimental
realizations of the target target dynamics.}
    \label{fig:IonQ}
\end{figure*}

Rather than implementing the conventional Strang decomposition explicitly
using half-time evolution operators, we retain the simpler gate structure
of a first-order decomposition while recovering second-order accuracy by
alternating the ordering of the two sectors between successive Trotter
steps. Specifically,
\begin{align}
U_A(\Delta t)
&=
e^{-iH_{X/XX}\Delta t}
e^{-iH_Z\Delta t},
\\
U_B(\Delta t)
&=
e^{-iH_Z\Delta t}
e^{-iH_{X/XX}\Delta t},
\end{align}
which are applied alternately throughout the simulation. Two successive
steps give
\begin{align}
U_B(\Delta t)U_A(\Delta t)
&=
e^{-iH_Z\Delta t}
e^{-iH_{X/XX}\Delta t}
e^{-iH_{X/XX}\Delta t}
e^{-iH_Z\Delta t}
\nonumber\\
&=
e^{-iH_Z\Delta t}
e^{-i2H_{X/XX}\Delta t}
e^{-iH_Z\Delta t},
\end{align}
where the second equality follows from the commutativity of the
$H_{X/XX}$ sector. This is precisely the Strang decomposition for an
evolution over $2\Delta t$, giving an overall second-order approximation
with local error $\mathcal{O}(\Delta t^3)$. The same construction is used
for both the truss gadget and the target $XX$ implementations, differing only in the
choice of $H_{X/XX}$.

The experiments are performed on the IonQ Forte trapped-ion quantum
processor, which provides 36 physical qubits with all-to-all connectivity
\cite{IonQForte}. Only five physical qubits are required here, so all
nearest- and next-nearest-neighbour interactions can be implemented
without connectivity overhead.

The resulting dynamics for the truss implementation are shown in
Figs.~\ref{fig:singleprob}(c) and \ref{fig:singleprob}(d), which display
the occupation probabilities of two arbitrarily chosen computational-basis states
$|00011\rangle$ and $|01010\rangle$, respectively. The IonQ measurements
are compared with the exact continuous-time evolution. The close agreement
over the full evolution window demonstrates that the truss gadget reproduces
the target dynamics at the level of individual many-body basis
states.

A complete comparison is presented in Fig.~\ref{fig:IonQ}, which shows the
time evolution of the probability distribution over all computational-basis
states for (a) the exact continuous-time evolution, (b) the truss
implementation on IonQ Forte, (c) the noiseless digital simulation of the
target $XX$ Hamiltonian, and (d) its direct implementation on IonQ Forte.
The noiseless $XX$ simulation in (c) closely reproduces the exact evolution
in (a), confirming the accuracy of the chosen Trotterization. The truss
implementation in (b) likewise captures the overall structure and temporal
evolution of the exact distribution, while (d) shows the corresponding
hardware realization using direct nearest-neighbour $XX$ interactions.

The difference between the two hardware implementations becomes apparent
when accounting for population outside the encoded subspace. For
the direct implementation of the target Hamiltonian using nearest-neighbour
$XX$ interactions, the logical Hilbert space is represented by an encoded
subspace (denoted by the subscript $\tilde{\mathcal{H}}$ throughout this section) of
the physical Hilbert space. In the absence of experimental errors, the
dynamics remain within this subspace. On hardware, however, errors can
populate states outside the encoded subspace. We therefore postselect the
measured probability distribution onto the encoded subspace and renormalize
it \cite{PhysRevLett.124.170501,Sarkar2026}. Denoting the corresponding
projector by $\mathcal{P}_{\tilde{\mathcal{H}}}$, the resulting probability of a
valid basis state $b$ is
\begin{equation}
P_b^{\tilde{\mathcal{H}}}(t)
=
\frac{
\langle b|
\mathcal{P}_{\tilde{\mathcal{H}}}\rho(t)\mathcal{P}_{\tilde{\mathcal{H}}}
|b\rangle
}{
\mathrm{Tr}\!\left[
\mathcal{P}_{\tilde{\mathcal{H}}}\rho(t)
\right]
}.
\end{equation}
The denominator renormalizes the distribution after discarding population
outside the even-parity subspace. This postselection discards approximately
$20\%$ of the measured population throughout the evolution window
($N_{\rm steps}=10$ at each evolution time), substantially reducing the
available experimental statistics. The resulting postselected distribution
nevertheless remains in good agreement with the corresponding noiseless
simulations.

The truss implementation does not require this postselection, since its
evolution takes place without imposed constraints on the physical Hilbert space. The
probability distribution is reconstructed after measurement through the
parity transformation
$\tilde{Z}_i=\hat{Z}_i\hat{Z}_{i+1}$. Every physical computational-basis
bitstring therefore has a well-defined image under the parity map. The distribution in the basis of the target Hamiltonian is obtained
by summing the probabilities of physical states that map to the same target-Hamiltonian basis-state, without discarding or renormalizing any population. Thus,
while the direct $XX$ implementation requires removal of population outside
the encoded subspace, the truss construction retains all measured outcomes
in reconstructing the target dynamics.

The agreement observed in Figs.~\ref{fig:singleprob} and
\ref{fig:IonQ} can be quantified over the complete many-body probability
distribution using the total variation distance,
\begin{equation}
D_{\mathrm{TV}}(P,Q)
=
\frac{1}{2}\sum_b
\left|P_b-Q_b\right|,
\end{equation}
where $P_b$ and $Q_b$ denote the probabilities of the computational-basis
state $|b\rangle$ in the two distributions being compared. For the noiseless
second-order Trotter evolution, the time-averaged total variation distance
from the exact continuous-time evolution is
$\overline{D}_{\mathrm{TV}}\simeq 0.04$, with a standard deviation of
approximately $0.04$ over the evolution window. The corresponding value for
the truss implementation on the IonQ Forte processor is
$\overline{D}_{\mathrm{TV}}\simeq 0.095$, with a standard deviation of
$0.02$. The larger deviation on hardware reflects the additional effects of
finite gate and measurement errors, while the larger spread of the Trotter
distance results from the increasing $\Delta t=T/N_{\mathrm{steps}}$ at
longer evolution times.

As a complementary measure, we also evaluate the Kullback--Leibler divergence,
\begin{equation}
D_{\mathrm{KL}}(P\Vert Q)
=
\sum_b P_b
\log\left(\frac{P_b}{Q_b}\right),
\end{equation}
between the full probability distributions. The corresponding time-averaged
values are approximately $0.01$ for the noiseless Trotter evolution and
$0.05$ for the truss implementation on IonQ. Both metrics show that the
truss implementation remains in close agreement with the exact many-body
dynamics, with the additional deviation on hardware attributable to
experimental errors.

\subsection{Demonstration of ground-state fidelity enhancement on a superconducting quantum annealer 
\label{sec:D-Wave}}
As a second benchmark of the truss gadget, we investigate its application to quantum annealing on the D-Wave Advantage system. In quantum annealing, the system is initialized in the ground state of a transverse-field driver Hamiltonian,
    \begin{align}
        \hat{H}_d = -\sum_i \hat{X}_i ,
    \end{align}
and is subsequently evolved toward the diagonal problem Hamiltonian
    \begin{align}
        \hat{H}_p = \sum_i h_i \hat{Z}_i + \sum_{\langle i,j\rangle} J_{ij}\hat{Z}_i\hat{Z}_j.
    \end{align}
The local fields $h_i$ and two-body couplings $J_{ij}$ encode a combinatorial optimisation problem, whose solution is represented by a ground state of $\hat{H}_p$. The corresponding annealing Hamiltonian is
    \begin{align}
        \hat{H}_{\mathrm{QA}}(s) = \beta(s) \hat{H}_d + \gamma(s) \hat{H}_p,
    \end{align}
where $\beta,\gamma:[0,1]\rightarrow\mathbb{R}_{\geq 0}$ denote the annealing schedule and $s\in[0,1]$ is the normalized time. For simulations of the system dynamics that are directly compared with experimental data, we use the native annealing schedules of the D-Wave Advantage system~\cite{dwave_advantage2_system2}. Otherwise, we use the linear annealing schedule $\beta(s)=1-s$ and $\gamma(s)=s$.

In the ideal adiabatic limit, a sufficiently slow interpolation allows the system to remain close to the instantaneous ground state, resulting in a high probability of measuring the ground state of $\hat{H}_p$ at the end of the anneal. In practice, however, transitions out of the instantaneous ground state may occur near avoided level crossings, where the spectral gap between the ground state and an excited state becomes small. Such transitions can substantially reduce the probability of obtaining the solution. In this context, perturbative crossings \cite{Ami09} constitute a major bottleneck, as they are associated with gaps that decrease exponentially with system size and can induce transitions into competing states separated from the solution state by large Hamming distances.

In the following, we consider an illustrative example problem whose quantum annealing spectrum exhibits a perturbative crossing which can be softened by the presence of an appropriately placed $XX$ coupling. We demonstrate that the effective $XX$ interaction, implemented using the truss gadget, achieves the expected ground-state fidelity enhancement when implemented on the D-Wave Advantage2\_system2 QPU.

\subsubsection{Problem Instance 
\label{sec:D-Wave:example}}

\begin{figure*}[htbp]
    \centering
    \includegraphics[width=\linewidth]{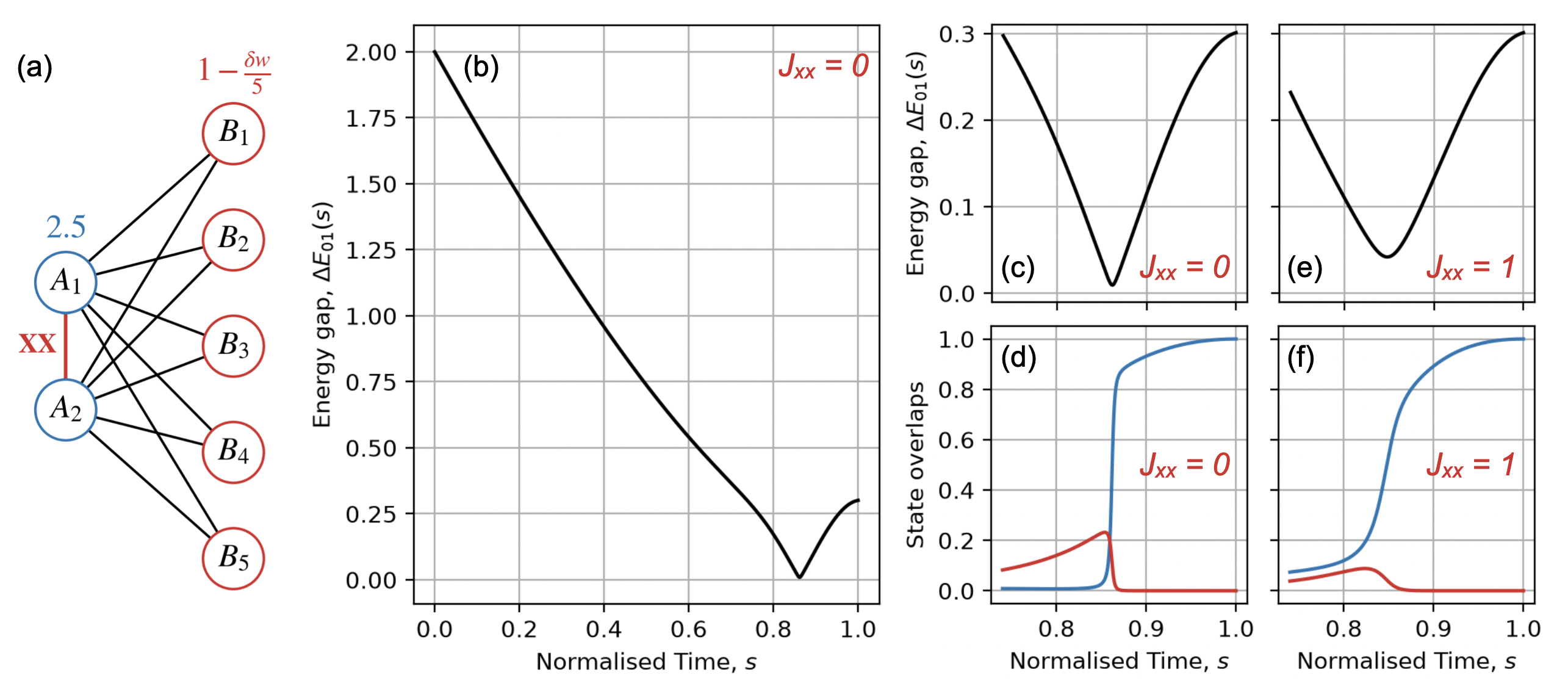}
    \caption{ (a) shows the $K_{2,5}$ MWIS graph utilised in section \ref{sec:D-Wave}. Panels (b-f) show numerical results associated with this graph with the hardness parameter set to $\delta w=1.4$, as obtained by exact diagonalisation. Panel (b) shows the evolution of the gap, $\Delta E_{01}(s)$, over the course of the anneal without the presence of an $XX$ interaction. Plots (c-f) show the evolution around the location of the perturbative crossing without and with the presence of an $XX$ interaction (between qubits $A_1$ and $A_2$) on the left and right respectively. Plots (c) and (e) show the evolution of the gap, $\Delta E_{01}(s)$. Plots (d) and (f) show the evolution of the instantaneous state overlaps $\abs{\braket{E_0(s)}{E_0}}$ (blue) and $\abs{\braket{E_0(s)}{E_1}}$ (red). }
    \label{fig:DWave:example}
\end{figure*}

The example problem that we consider is an instance of the maximum weighted independent set (MWIS) problem based on the instances used in~\cite{Fei24}. The problem, depicted in Fig.~\ref{fig:DWave:example}(a), has a $K_{2,5}$ bipartite structure. The two vertices in the first sub-graph, $A$, are each given a weight of $2.5$ while the five vertices in the second sub-graph, $B$, are given weights of $1-\delta w/5$. The result is a global optimum of weight $5$, consisting of the two vertices in $A$, and a local optimum of weight $5-\delta w$, consisting of the five vertices in $B$. The global and local optima are separated by a Hamming distance of seven. When encoded into an Ising Hamiltonian, this setup produces a problem spectrum with properties that result in the formation of a perturbative crossing.

By tuning $\delta w$, we can adjust the competitiveness of the local optimum. Smaller values of  $\delta w$ result in smaller gap minima associated with the perturbative crossing present in the anneal of the problem meaning that longer run-times are needed to achieve the same ground-state fidelity. In particular, when the problem graph is scaled to larger bi-partite systems, tuning this parameter results in quantitively different gap scaling such that we can consider $\delta w$ a hardness parameter for the problem \cite{Fei24}. We make use of this hardness parameter in section \ref{sec:D-Wave:implementation}. For now, we set $\delta w$ to $1.4$ such that the local optimum has a total weight of $3.6$.

The MWIS problem on a graph defined by the set of vertices, $V$, and edges $E$, can be encoded into a Hamiltonian of the form,
\begin{equation}
    \hat{H}_p = \sum \limits_{i \in V} (d_i J_{zz} - 2w_i)\hat{Z}_i +  J_{zz} \sum \limits_{(i,j) \in E} \hat{Z}_i \hat{Z}_j ,
    \label{eq:MWIS}
\end{equation}
where $w_i$ and $d_i$ are respectively the weight and degree  of vertex $i$, and $J_{zz}$ is the strength of the penalty enforcing the independent set condition. We take $J_{zz}=5$ and the fields and coupling strengths obtained from Eq.~\ref{eq:MWIS} are normalised to unity. 

The evolution of the energy gap between the ground and first excited state, $\Delta E_{01}(s)$, associated with annealing of the problem described by Fig.~\ref{fig:DWave:example}(a), but without an $XX$ interaction, is shown in Fig.~\ref{fig:DWave:example}(b,c). The expected perturbative crossing can be seen as a gap minimum at $s \approx 0.85$. Fig.~\ref{fig:DWave:example}(d) shows the evolution of the instantaneous ground-state overlap with the problem ground- and first-excited states in blue and red respectively around the location of the perturbative crossing. We see that, prior to the location of the gap minimum, there is an accumulation of the first-excited state overlap while the ground-state overlap remains approximately zero. This is followed by a sharp transition into the problem ground-state. For the system to remain in the instantaneous ground-state as the anneal moves through this point, the system must transition between states that are a Hamming distance $7$ apart.

The addition of an $XX$ interaction between qubits $A_1$ and $A_2$ is able to soften this perturbative crossing by helping to guide the anneal towards the global-minimum, thereby enhancing the minimum spectral gap and smoothing the transition. The placement of this coupling can be motivated through perturbation theory using similar arguments to those outlined in \cite{Fei24}. We introduce this $XX$ coupling to the annealing protocol as part of the driver such that the updated driver reads
\begin{equation}
    \hat{H}_d= -\left( \sum_{i\in V} \hat{X}_i + J_{xx} \hat{X}_{A_1}\hat{X}_{A_2} \right).
\end{equation}
Fig.~\ref{fig:DWave:example}(e) shows the enhanced gap minimum, which has been increased from $\Delta E^\text{min}_{01}(s)=0.0094$ to $0.0419$. In Fig.~\ref{fig:DWave:example}(f) we see that the accumulation of the problem first-excited-state is suppressed such that there is a smoother transition towards the global optimum.

\subsubsection{Implementation on Superconducting Hardware
\label{sec:D-Wave:implementation}}

To implement the $XX$ coupling between qubits $A_1$ and $A_2$, we replace these qubits by a truss gadget. The set of three qubits forming the truss are denoted by $T$, such that the logical $Z$ operators of the original problem graph are encoded as the parities
    \begin{align}
        \tilde{Z}_{A_1} &\equiv \hat{Z}_{T_1} \hat{Z}_{T_3},\\
        \tilde{Z}_{A_2} &\equiv \hat{Z}_{T_2} \hat{Z}_{T_3}.
    \end{align}
The effective $XX$ coupling between $A_1$ and $A_2$, as well as their local $Z$ fields, are realized as described in section \ref{sec:gadget}.

In the original $K_{2,5}$ graph, each logical spin in $A$ is coupled via a $ZZ$ interaction to every spin in $B$. Because these interactions couple spins inside the truss to spins outside it, the truss encoding raises these interactions to three-body $ZZZ$ couplings of the form
    \begin{align}
        \tilde{Z}_{A_i} \tilde{Z}_{B_j} = \hat{Z}_{T_i} \hat{Z}_{T_3} \hat{Z}_{B_j},
    \end{align}
where $i\in \{1,2\}$ and $j\in \{1,2,3,4,5\}$. Since three-body interactions cannot be implemented directly on the D-Wave annealer, we introduce additional auxiliary variables $C_k$ to reduce the interactions to two-body terms
    \begin{multline}
         \hat{Z}_{T_i} \hat{Z}_{T_3} \hat{Z}_{B_j} \to \left( \hat{Z}_{T_i} \hat{Z}_{T_3} + \hat{Z}_{T_3} \hat{Z}_{B_j} + \hat{Z}_{T_i} \hat{Z}_{B_j} \right) \\ +  \left( \hat{Z}_{T_i} + \hat{Z}_{T_3} + \hat{Z}_{B_j} \right) \\ + 2\left( \hat{Z}_{C_k} \hat{Z}_{T_i} + \hat{Z}_{C_k} \hat{Z}_{T_3} + \hat{Z}_{C_k} \hat{Z}_{B_j} \right) + 2\hat{Z}_{C_k}.
        \label{eq:ZZZ}
    \end{multline}
Note that Eq.~\ref{eq:ZZZ} preserves the relevant energy-level structure of the MWIS instance~\cite{Leib_2016}, thus keeping the perturbative level crossing in the instantaneous spectrum. Since each logical  coupling between $A_i$ and $B_j$ introduces one additional auxiliary qubit, the truss-encoded $K_{2,5}$ MWIS instance requires $18$ physical qubits in total. Its encoded Ising graph is shown in Fig.~\ref{fig:DWave:samples}(b). A complete expression for the resulting Hamiltonian is provided in Appendix~\ref{app:encoded-hamiltonian}.

To run the encoded MWIS instance on the D-Wave Advantage2\_system2 QPU~\cite{dwave_advantage2_system2}, an additional minor-embedding step is required to map the encoded Ising graph onto the hardware graph. We use the \texttt{minorminer} tool provided by the D-Wave Ocean SDK~\cite{dwave_ocean} to generate random hardware embeddings requiring between $20$ and $27$ physical qubits. Because minor embedding represents individual problem variables by chains of physical qubits, the resulting embedding can influence the annealing dynamics, particularly in the vicinity of the perturbative crossing. To reduce embedding-specific effects, we partition the QPU into an $8\times8$ grid and generate an independent random embedding within each of the resulting $64$ cells (see Fig.~\ref{fig:DWave:embedding} in Appendix~\ref{app:encoded-hamiltonian}). Distributing the embeddings across the QPU additionally allows us to average over spatial variations in noise, calibration, and hardware performance. To further mitigate calibration errors and systematic biases, we apply spin-reversal transformations and enable automatic scaling of the local fields and couplings.

\begin{figure}[htbp]
    \centering
    \includegraphics[width=\linewidth]{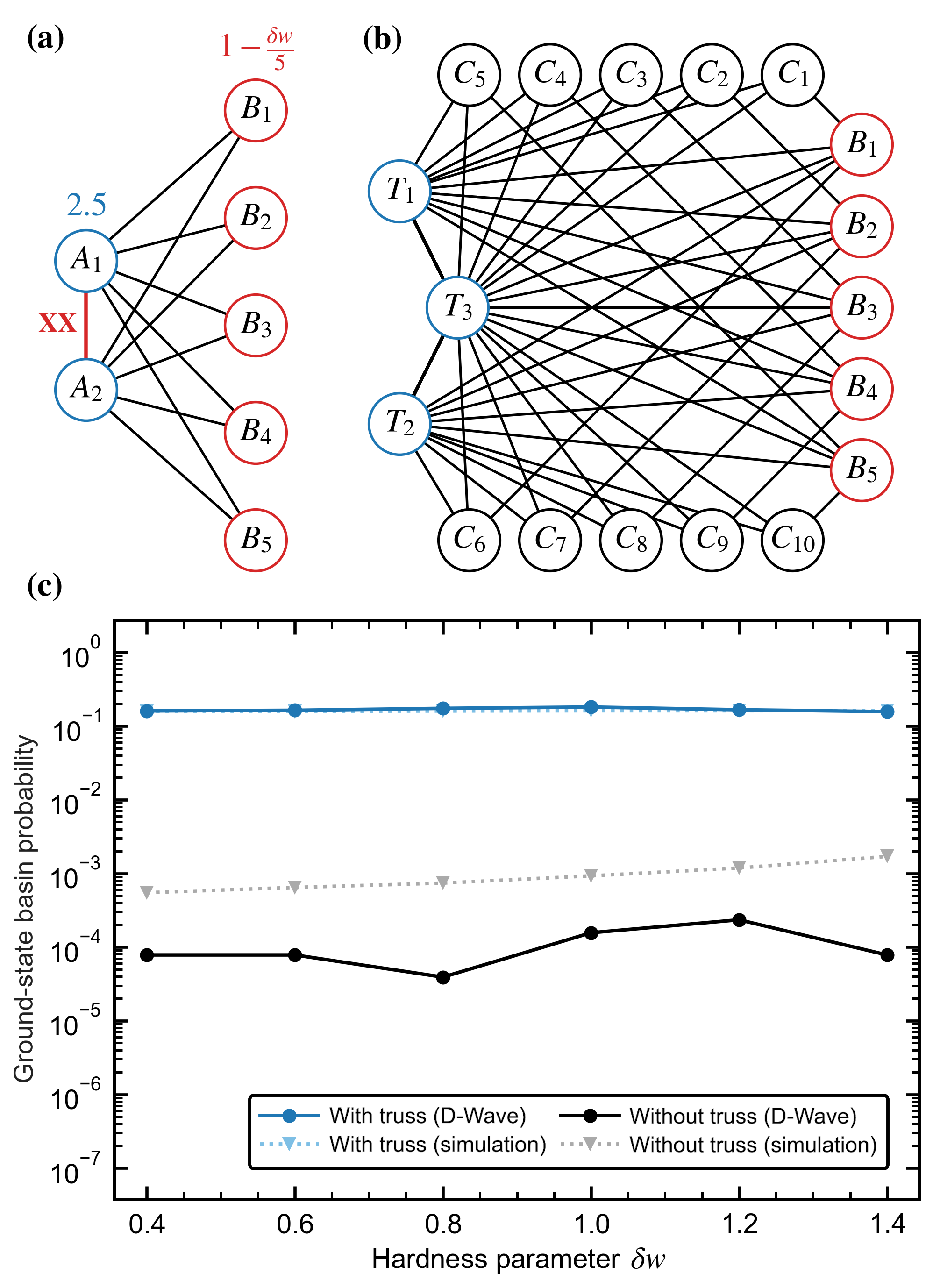}
    \caption{Performance of a $K_{2,5}$ MWIS instance using the truss gadget on the D-Wave Advantage2\_system2. (a) Logical MWIS graph with an $XX$ coupling between qubits $A_1$ and $A_2$. (b) Truss encoding of the $XX$ coupling in terms of native $X$ and $Z$ fields and $ZZ$ interactions. The variables $C_{1\text{--}10}$ denote auxiliary variables required to implement the resulting $ZZZ$ couplings. (c) Global-minimum basin probability as a function of the hardness parameter $\delta w$, comparing implementations with $XX$ coupling ($20$--$27$ physical qubits) and without $XX$ coupling ($7$ physical qubits). Solid lines show experimental results obtained on the D-Wave Advantage2\_system2 using $64$ simultaneous minor embeddings (see Fig.~\ref{fig:DWave:embedding}) and $10{,}000$ QPU samples per embedding, corresponding to $6.4\times10^{5}$ measurements in total, with an annealing time of $0.5\,\mu\mathrm{s}$. Auto-scaling and spin-reversal transformations are applied. Dotted lines show results from ideal state-vector simulations in the absence of decoherence. The corresponding ground-state probabilities are shown in Fig.~\ref{fig:DWave:samples_gs}.}
    \label{fig:DWave:samples}
\end{figure}

Fig.~\ref{fig:DWave:samples} shows the probability of the system ending in the global-minimum basin, defined as the total probability of sampling a computational-basis state within Hamming distance two of the solution state. Results are shown both with the truss gadget (solid red) and without it (solid black) as a function of the hardness parameter $\delta w$. We use this basin probability rather than the exact ground-state probability because it provides a more robust measure of successful passage through the perturbative crossing. In a finite-temperature system, probability mass may be distributed among low-lying configurations near a classical minimum rather than being concentrated entirely in the exact ground state. Consequently, after the system has successfully traversed the perturbative crossing and entered the global-minimum basin, subsequent intra-basin excitations can reduce the exact ground-state probability without indicating a failure to pass the perturbative crossing. By summing over configurations within Hamming distance two of the global minimum, the basin probability captures such nearby excitations while remaining sensitive to population transfers into the competing local minimum. We therefore use the basin probability as our primary observable and report the corresponding ground-state probabilities in Appendix~\ref{app:DWave:samples_gs} as a robustness check. Both observables exhibit the same qualitative enhancement in the presence of the truss gadget, indicating that our conclusions are insensitive to the choice of measure.

The results in Fig.~\ref{fig:DWave:samples} show that incorporating the truss gadget can increase the probability of reaching the global-minimum basin by more than three orders of magnitude. For example, at $\delta w=0.4$, the probability increases from $7.8\times10^{-5}$ to $1.6\times10^{-1}$. This improvement is particularly notable given the additional hardware overhead associated with the truss implementation. First, after minor embedding, the truss-encoded instance requires $20$--$27$ physical qubits compared with only $7$ physical qubits for the native $K_{2,5}$ MWIS instance. In particular, the introduction of qubit chains creates additional low-energy states within the broader local- and global-minimum basins. These additional states can trap probability mass away from the respective logical minimum. Second, reducing the logical $ZZZ$ interactions to two-local terms substantially increases the scale of the required physical fields and couplings. To keep all parameters within the programmable ranges of the D-Wave Advantage QPU, the resulting Hamiltonian must therefore be rescaled by a factor of approximately $1/18$. This rescaling reduces the overall problem energy scale relative to the device temperature and environmental noise. Consequently, even after successfully traversing the perturbative crossing into the global-minimum basin, noise-induced transitions can redistribute probability mass away from the global minimum and, in principle, back toward competing low-energy states. Despite these hardware limitations, the truss implementation produces a substantial enhancement in the global-minimum basin probability across all investigated hardness values. This result suggests that hardware supporting native $ZZZ$ interactions, and thereby avoiding both the auxiliary-qubit overhead and the associated reduction in energy scale, could enable further enhancement.

The experimental results are further supported by ideal state-vector simulations of the corresponding minor-embedded problem instances (dashed curves in Fig.~\ref{fig:DWave:samples}). For the truss-encoded instance, we observe excellent agreement between experiment and simulation. By contrast, the native instance exhibits an offset of approximately one order of magnitude, with the simulated probabilities exceeding the experimentally measured values.

In summary, the close agreement between numerical and experimental results for the encoded problem demonstrates that the truss gadget provides a viable approach for realizing effective $XX$ couplings on current quantum-annealing hardware. Moreover, comparison with the native problem instance shows that the gadget can substantially enhance the probability of reaching the global-minimum basin despite the additional embedding and energy-scale overhead. These results demonstrate that parity-based implementations of nonstoquastic interactions can improve the performance of quantum annealing for problem instances dominated by perturbative crossings.

\section{ Conclusion \label{sec:conclusion} }

We have shown that the truss gadget enables the implementation of a logical $XX$ interaction using only physical $X$ and $Z$ fields and $ZZ$ interactions with minimal hardware overhead. We have experimentally demonstrated the accuracy and utility of the truss gadget using both an ion-trap gate-based quantum computer and a superconducting quantum annealer. 

In general terms, $XX$ interactions in quantum annealing can be implemented either directly by engineering a novel hardware element or indirectly by combining existing hardware elements in a novel way. The truss gadget belongs to the latter category. Unlike perturbative or asymptotic constructions, however, the truss interaction is exact by construction. 
In the context of superconducting flux qubits of the type used in the D-Wave annealer, direct hardware implementation of an $XX$ interaction could in principle be achieved using either an inductive \cite{zhang2023tunableinductivecouplerhigh} or a capacitive \cite{PhysRevApplied.13.034037} coupler. Kerman \cite{Kerman_2019} has however shown that in both cases a non-negligible $XX$ interaction can only be achieved if the local $X$ fields applied to both coupled qubits are large. 

The truss construction is also compatible in principle with neutral-atom platforms, where the required longitudinal interactions arise naturally from the distance-dependent interactions between Rydberg-excited atoms. Implementing the truss geometry on such platforms requires careful engineering of the atomic positions, due to the long-range nature of the interactions \cite{Hollerith2022,Sarkar2025}. Determining geometries that realise the required interaction graph while suppressing or incorporating unwanted couplings is an interesting direction for future work.

Apart from applications in quantum annealing, the truss gadget provides a direct construction for realizing non-native interacting spin Hamiltonians from a restricted set of hardware interactions. This fungibility of physical realizations provides a useful route for cross-validating quantum simulations: the same effective Hamiltonian can be implemented through different physical constructions or encodings, which must produce identical dynamics in the ideal limit. Combining the generated \(XX\) interactions with \(ZZ\) couplings and longitudinal \(Z\) fields gives access to anisotropic spin models beyond the transverse-field Ising model, closely related to the extensively studied \(XXZ\) and \(XYZ\) families \cite{Trebst2022,Luo2025}. The absence of a \(YY\) term corresponds to a particular anisotropic limit of the \(XYZ\) model; generating independent \(YY\) interactions would require an additional phase degree of freedom in the local transverse drive. More broadly, proposals for engineering generic \(QQ\) interactions, with \(Q\in\{X,Y,Z\}\), using superconducting phase-slip devices point toward routes for extending this approach to more general spin Hamiltonians \cite{Kerman_2019}. The same parity-based construction also suggests applications to gate-based quantum processors. With phase control of the local transverse drive, the gadget could in principle be extended to generate \(XX+YY\) interactions, enabling particle-number-conserving exchange dynamics. The encoding offers an advantage: while two-qubit gate errors in a direct encoded implementation can populate states outside the logical subspace and require postselection, every physical bit string produced by the truss gadget setup has a well-defined logical image, allowing all measurement outcomes to be retained.

While the truss construction therefore has broader applications in Hamiltonian engineering and quantum simulation, our main focus here is its use for enhancing the minimum spectral gap in quantum annealing. Our experimental demonstrations involve two small-scale Hamiltonians: a ring of five logical qubits and a weighted maximum-independent-set problem on the seven-vertex bipartite \(K_{2,5}\) graph. The gap-enhancement mechanism is however certainly not specific to these problem classes. It has previously been shown numerically that the addition of $XX$ interactions can enhance the quantum annealing gap minimum for randomly-generated MWIS instances on Erd\H{o}s-R\'{e}nyi and Barab\'{a}si-Albert graphs \cite{nut24}, while analytical results have established gap enhancement in the thermodynamic limit (\textit{i.e.} at infinite system size) for the weak-strong cluster problem~\cite{Tak20}. However, the one-dimensional structure of the truss does impose a constraint on how these interactions can be embedded: the logical $XX$ couplings must form non-intersecting one-dimensional chains. When the objective is to maximise the number of logical vertex pairs coupled by $XX$ interactions while maintaining local $X$ interactions, the corresponding distribution of truss gadgets with three physical qubits can be determined in polynomial time using Edmonds's blossom algorithm~\cite{Edmonds_1965}. 

An important consequence of our construction is that the logical $XX$ coupling always has the same sign as the physical $X$ field, as shown in Fig.~\ref{fig:truss}. Hence the logical Hamiltonian is necessarily stoquastic for homogeneous driving. This restricts the construction to stoquastic Hamiltonians and does not provide access to the non-stoquastic \(XX\) catalysts considered in some approaches to quantum annealing. Nevertheless, both analytical and numerical studies have demonstrated that stoquastic $XX$ interactions can enhance the minimum gap for a range of problem classes~\cite{Hormozi_2017,Mehta_2021,nut24,Crosson2020}.

As noted in section~\ref{sec:gadget}, third-order $ZZZ$ and fourth-order $ZZZZ$ physical interactions are required to implement logical $ZZ$ interactions between a qubit within the truss and (a) an unencoded qubit or (b) non-adjacent encoded qubits, respectively. While such terms are not directly available on current commercial hardware, several superconducting architectures have been proposed for implementing higher-order longitudinal interactions \cite{Chancellor2017CircuitDesign,PhysRevLett.129.220501,melanson2019tunablethreebodycouplersuperconducting}. Alternatively, standard ``HUBO-to-QUBO'' reduction techniques can be used to replace these higher-order terms by quadratic interactions, at the cost of a number of additional physical qubits that scales linearly with the interaction order \cite{dattani2019quadratizationdiscreteoptimizationquantum}. Thus, the remaining hardware requirements of the truss are compatible with existing routes for engineering higher-order longitudinal interactions or, alternatively, with standard reduction techniques.

Several other approaches have been proposed to enhance the minimum gap in quantum annealing, including exact \cite{PhysRevLett.109.115703} and approximate \cite{PhysRevLett.123.090602} counterdiabatic driving, diagonal catalysts \cite{PhysRevA.103.022608,cáliz2026reshapingquantumannealinglandscapes}, and inhomogeneous $X$ drivers \cite{PhysRevA.85.032303,PhysRevA.96.042322}. These approaches modify the annealing Hamiltonian in fundamentally different ways and involve distinct trade-offs between the interactions that can be generated, their implementation accuracy, and the required hardware resources. The truss gadget provides a complementary route: rather than engineering a new hardware element or modifying the annealing schedule, it converts a set of simple local interactions into an exact logical $XX$ interaction. In this sense, the truss gadget provides a general strategy for extending the range of Hamiltonians which are accessible to quantum-annealing hardware, while retaining the simplicity of the underlying interaction set.

The concept behind the truss gadget can be extended to implement general $k^\text{th}$ order transverse interactions with a $k+1$ qubit construction. This is achieved by having a single qubit shared between $k$ pairs whose parities encode the $k$ logical qubits such that local $X$ driving of the shared qubit effectively flips all the logical spins at once. Such higher order constructions do not come with any further demands on the order of the required physical $Z$ terms. The extension of such higher order gadgets to encode multiple, linked, higher order $X$ interactions is however less straightforward than the truss-chain construction.

\begin{acknowledgments}
The authors gratefully acknowledge the Gauss Centre for Supercomputing e.V. (www.gauss-centre.eu) for funding this project by providing computing time on the GCS Supercomputer JUWELS~\cite{Juwels_Cluster} at Jülich Supercomputing Centre (JSC). The authors gratefully acknowledge the Jülich Supercomputing Centre (https://www.fz-juelich.de/ias/jsc) for funding this project by providing computing time on the D-Wave Advantage™ System JUPSI through the Jülich UNified Infrastructure for Quantum computing (JUNIQ). This project was also funded and supported by the UK National Quantum Computer Centre [NQCC200921], which is a UKRI Centre and part of the UK National Quantum Technologies Programme (NQTP)

RB, NF, SB and PAW were funded by EPSRC (grant ref. EP/Y004590/1). NF and PAW were additionally supported by the QCi3 Hub, funded by EPSRC (grant ref. EP/Z53318X/1) 
\end{acknowledgments}

\section*{Author Contributions}
Truss gadget conception: RJB. Ising problem construction and simulation: NF, SS. Mapping onto ion-trap hardware: RG. Experiments on ion-trap hardware: RG and NF. Mapping onto superconducting hardware: SS, NF and PAW. Experiments on superconducting hardware: SS and KM. Data analysis and interpretation: NF, RG, SS, SB and PAW. Manuscript preparation: RJB, NF, RG, SS, SB and PAW. Theoretical project leadership and supervision: SB. Experimental project leadership and supervision: PAW.

\bibliography{refs.bib}

\appendix

\section{Truss gadget applied to a loop \label{app:loop_details} }

This appendix details how the truss gadget leads to the target Hamiltonian in section~\ref{sec:IonQ}. To the physical Hamiltonian describing the truss gadget (Eq.~\ref{eq:H_truss_phy}) we introduce nearest and next-nearest $\hat{Z}\hat{Z}$ couplings between physical qubits at each end of the gadget, forming a closed loop. The resulting Hamiltonian is
\begin{multline}
    \hat{H}_\text{Phys}^\text{Truss}=-\sum_{i=1}^{n} d_i \hat{X}_i 
    +\sum_{i=1}^{n-1} h_{i} \hat{Z}_{i} \hat{Z}_{i+1}
    +\sum_{i=1}^{n-2} J_{i,i+1} \hat{Z}_{i} \hat{Z}_{i+2} \\
    +K_{1,n}\hat{Z}_1\hat{Z}_n+K_{1,n-1}\hat{Z}_1\hat{Z}_{n-1}+K_{2,n}\hat{Z}_2\hat{Z}_{n},
    \label{eq:H_truss_phy_loop}
\end{multline}
where $K_{i,j}$ denotes the strength of the introduced coupling. Through application of Eq.~\ref{eq:cong_z}, the resulting logical Hamiltonian in this case is
\begin{multline}
    \tilde{H}_\text{Logical}^\text{Truss}=-\sum_{i=1}^{n-2} d_{i+1} \tilde{X}_i\tilde{X}_{i+1}-d_1 \tilde{X}_1-d_{n} \tilde{X}_{n-1}\\
    +\sum_{i=1}^{n-1} h_i \tilde{Z}_i+\sum_{i=1}^{n-2} J_{i,i+1} \tilde{Z}_i\tilde{Z}_{i+1}\\+K_{1,n}\prod_{i=1}^{n-1} \tilde{Z}_i+K_{1,n-1}\prod_{i=1}^{n-2} \tilde{Z}_i+K_{2,n} \prod_{i=2}^{n-1} \tilde{Z}_i.
    \label{eq:H_truss_log_loop}
\end{multline}
As in section.~\ref{sec:gadget}, we have gone from $n$ physical qubits to $n-1$ logical qubits. 

It remains to show that this logical Hamiltonian is equivalent to the target Hamiltonian in section~\ref{sec:IonQ}. To this end we introduce a second encoding utilizing an additional global-parity qubit. We use $\bar{X}$ and $\bar{Z}$ to denote the Pauli operators in this new global-parity qubit encoding. Each of the original logical $Z$ operators correspond directly to $Z$ operators in the global-parity qubit encoding such that
\begin{equation}
    \tilde{Z}_i \equiv \bar{Z}_i, \text{ for all }\, 1\leq i \leq n-1.
\end{equation}
The global-parity qubit $\bar{Z}_{n}$ is introduced with the global-constraint
\begin{equation}
    \prod_{i=1}^n\bar{Z}_i = (-1)^{\mathcal{P}}
\end{equation}
where $\mathcal{P}\in \{0,1\}$ denotes the parity of valid computational basis states. Therefore, the value of $\bar{Z}_n$ is fixed by the parity of the $n-1$ other qubits. Hence, the $n^\text{th}$ qubit plays the role of a classical parity bit, which can be used to detect an erroneous single bit-flip \cite{Hamming_1950}. It follows that the terms in the logical Hamiltonian introduced by coupling the ends of the truss gadget can be expressed as
\begin{align}
    \prod_{i=1}^{n-1} \tilde{Z}_i & \equiv \bar{Z}_n \prod_{i=1}^{n} \bar{Z}_i =(-1)^{\mathcal{P}}\bar{Z}_n \\
    \prod_{i=1}^{n-2} \tilde{Z}_i & \equiv \bar{Z}_{n-1}\bar{Z}_n \prod_{i=1}^{n} \bar{Z}_i = (-1)^{\mathcal{P}}\bar{Z}_{n-1}\bar{Z}_n\\
    \prod_{i=2}^{n-1} \tilde{Z}_i & \equiv \bar{Z}_{1}\bar{Z}_n \prod_{i=1}^{n} \bar{Z}_i = (-1)^{\mathcal{P}}\bar{Z}_{1}\bar{Z}_n.
\end{align}
The transverse-field is encoded as
\begin{equation}
    \tilde{X}_i \equiv \bar{X}_i\bar{X}_n.
\end{equation}
Using the global-parity qubit encoding the logical Hamiltonian in Eq.~\ref{eq:H_truss_log_loop} becomes
\begin{multline}
    \bar{H}_\text{Target}^\text{Global}=-\sum_{i=1}^{n-2} d_{i+1} \bar{X}_i\bar{X}_{i+1}-d_1 \bar{X}_1\bar{X}_n - d_{n} \bar{X}_{n-1}\bar{X}_n\\
    +\sum_{i=1}^{n-1} h_i \bar{Z}_i+\sum_{i=1}^{n-2} J_{i,i+1} \bar{Z}_i\bar{Z}_{i+1}\\+(-1)^{\mathcal{P}}\left(K_{1,n}\bar{Z}_n+K_{1,n-1}\bar{Z}_{n-1}\bar{Z}_n+K_{2,n}\bar{Z}_{2}\bar{Z}_n\right).
    \label{eq:H_global_log_loop}
\end{multline}
with the constraint $\prod_{i=1}^n\bar{Z}_i = (-1)^{\mathcal{P}}$. As with the truss gadget, $n-1$ logical qubits have been encoded in $n$ physical qubits. We reserve the name of logical Hamiltonian for Eq.~\ref{eq:H_truss_log_loop}, since this Hamiltonian acts on logical qubits \textit{i.e.} qubits each with their own independent degree of freedom. 

The global-parity qubit encoding has the benefit of reducing the support of the interactions in the logical Hamiltonian, while providing a physically distinct implementation for comparison to the truss gadget. One can formulate a direct equivalence between the $Z$ values for each of the $n$ qubits in both the truss and global-parity qubit gadget as
\begin{equation}
    \bar{Z}_i \equiv \begin{cases} \hat{Z}_i \hat{Z}_{i+1}\, &\text{ if } 1\leq i \leq n-1 \\
    (-1)^{\mathcal{P}}\hat{Z}_1 \hat{Z}_{n}\, &\text{ if } i = n
    \end{cases}
    \label{eq:loop_decoder}
\end{equation}
In section~\ref{sec:IonQ}, the parity is fixed to the even-parity sector $\mathcal{P} = 0$ and decoded through Eq.~\ref{eq:loop_decoder}. This constraint is enforced through the choice of initial state and by postselecting on even-parity states in the presence of experimental errors. 

In this appendix we have shown how to recover the Hamiltonian, illustrated in Fig.~\ref{fig:singleprob}(b) from the truss gadget with the additional $\hat{Z}\hat{Z}$ terms depicted in Fig.~\ref{fig:singleprob}(a). One can also show the opposite direction, where any one of the qubits in Fig.~\ref{fig:singleprob}(b) can be chosen to be the global-parity qubit.

\section{ Encoded Hamiltonian  \label{app:encoded-hamiltonian} }

Starting from Eq.~\ref{eq:MWIS}, the logical Hamiltonian for the $K_{2,5}$ bipartite MWIS instance depicted in Fig.~\ref{fig:DWave:samples}(a) is given by
\begin{multline}
    \tilde{H}_p = h_A (\tilde{Z}_{A_1}  +\tilde{Z}_{A_2}) + h_B \sum \limits_{i=1}^{5} \tilde{Z}_{B_i}  \\ + J_{zz} (\tilde{Z}_{A_1}  +\tilde{Z}_{A_2}) \sum \limits_{i =1}^{5} \tilde{Z}_{B_i} 
\end{multline}
with $h_A=2\times J_{zz} - 2 \times2.5=20$ and $h_B=10\times J_{zz} - 2 (1-\delta/5)=8-2\delta/5$ corresponding to node weights of $2.5$ for nodes in $A$ and weights of $1-\delta/5$ for nodes in $B$, with edge penalty, $J_{zz}=5$.

Replacing the nodes in $A$ with the truss, $T_i$ for $i\in\{1,2,3\}$, the problem Hamiltonian becomes,
\begin{multline}
    \hat{H}^\text{Truss}_p = h_A (\hat{Z}_{T_1}  +\hat{Z}_{T_2})\hat{Z}_{T_3}+ h_B \sum \limits_{i=1}^5 \hat{Z}_{B_i} \\  + J_{zz}  (\hat{Z}_{T_1}  +\hat{Z}_{T_2})\hat{Z}_{T_3} \sum \limits_{i = 1}^{5} \hat{Z}_{B_i}.
\end{multline}
Each of the ten $ZZZ$ interactions in this Hamiltonian need to be replaced by the two-local Hamiltonian given by Eq.~\ref{eq:ZZZ}. This introduces a new auxiliary qubits per three-body interaction. The ordered set of auxiliary qubits is labelled by $C$ with the $i^\text{th}$ auxiliary qubit denoted as $C_i$. The final Hamiltonian, with the truss gadget and auxiliary qubit implementation of the $ZZZ$ interactions, is then
\begin{multline}
    \hat{H}^\text{final}_p = \left(h_A + 5 J_{zz}\right)\left(\hat{Z}_{T_1}  +\hat{Z}_{T_2}\right)\hat{Z}_{T_3}\\
    +  5 J_{zz} \left(\hat{Z}_{T_1} + \hat{Z}_{T_2} + 2\hat{Z}_{T_3}\right) + 2J_{zz} \left(1+\hat{Z}_{T_3}\right) \sum \limits_{i=1}^{10} \hat{Z}_{C_i} \\ 
    + \left[ h_B+J_{zz} \left( 2+ \hat{Z}_{T_1} + \hat{Z}_{T_2} + 2\hat{Z}_{T_3} \right)\right]
    \sum \limits_{i=1}^5 \hat{Z}_{B_i} \\
    + 2 J_{zz} \sum \limits_{i=1}^{5} 
    \left[ \left( \hat{Z}_{T_1} + \hat{Z}_{B_i}\right) \hat{Z}_{C_{i}} + \left( \hat{Z}_{T_2} + \hat{Z}_{B_i} \right) \hat{Z}_{C_{5+i}} \right].
    \label{eq:app:h_final}
\end{multline}

The resultant physical graph has $18$ qubits. The shared qubit in the truss, $T_3$, is coupled to every other qubit such that it has degree $17$. The remaining two truss qubits, $T_1$ and $T_2$, have degree $11$ each, the vertices in $B$ have degree $5$, and the auxiliary qubits in $C$ each have degree $3$.

\begin{figure}[htbp]
    \centering
    \includegraphics[width=\linewidth]{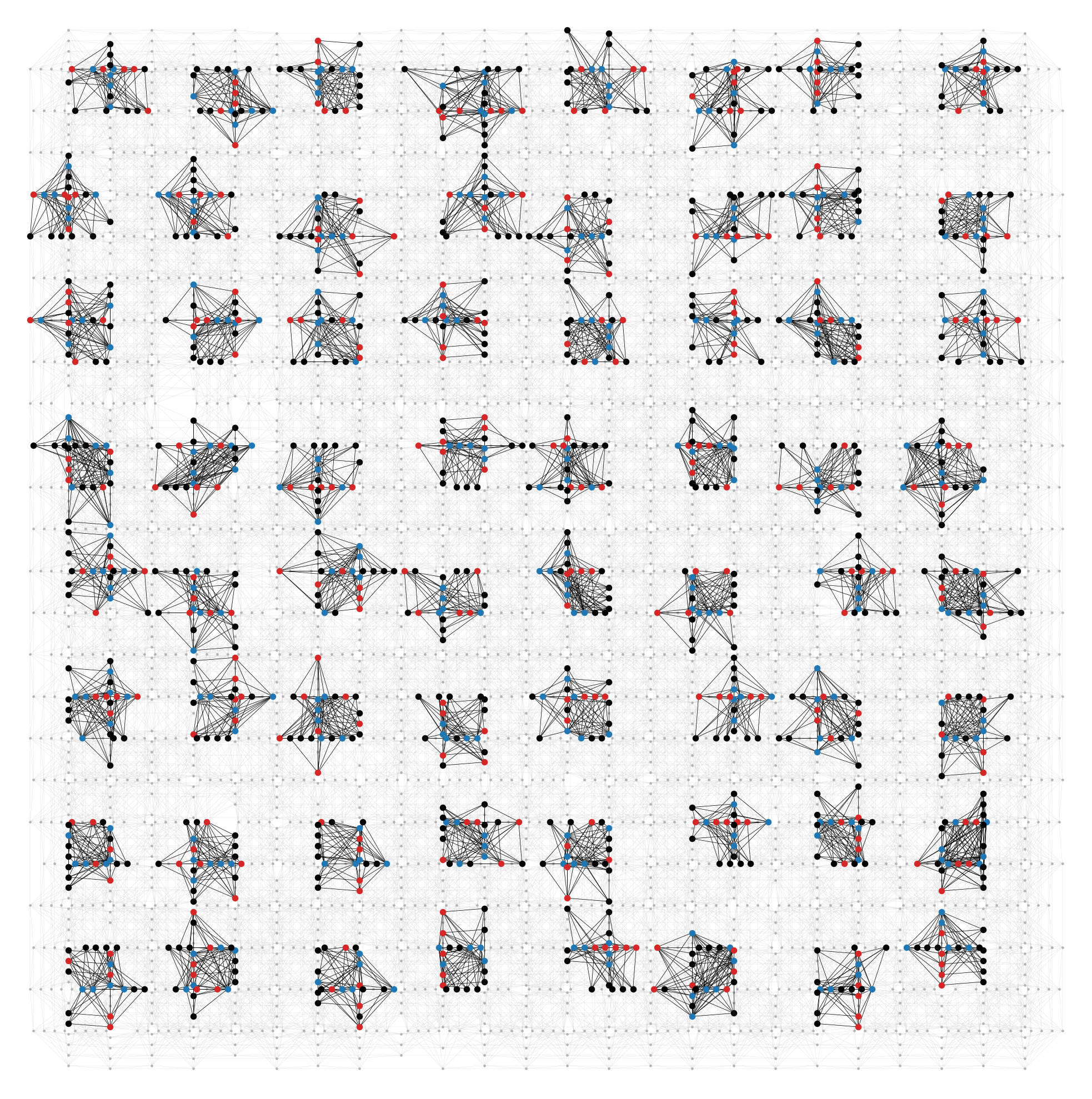}
    \caption{Minor embedding of the truss-encoded $K_{2,5}$ bi-partite MWIS instance on the Advantage2\_system2 QPU using between $20$ and $27$ physical qubits. A total of $64$ independent minor embeddings are distributed across an $8\times8$ grid. Node colours follow the encoding convention of Fig.~\ref{fig:DWave:samples_gs}(b): blue nodes denote the truss variables $T$, red nodes denote the local-minimum variables $B$, and black nodes correspond to the auxiliary qubits $C$.}
    \label{fig:DWave:embedding}
\end{figure}

In order to implement the encoded problem Hamiltonian $\hat{H}^\text{final}_p$ on the D-Wave Advantage2\_system2 QPU~\cite{dwave_advantage2_system2}, an additional minor-embedding step is required to map the logical couplings in Eq.~\ref{eq:app:h_final} onto the hardware graph. To this end, we use the \texttt{minorminer} package from the D-Wave Ocean SDK~\cite{dwave_ocean}. The resulting minor embeddings require between $20$ and $27$ physical qubits, with logical variables represented by chains of physical qubits. Since these qubit chains can modify the effective annealing dynamics, particularly near the perturbative crossing, we average over a large ensemble of independently generated embeddings. Specifically, the QPU is divided into an $8\times8$ array of qubit cells, with one random embedding generated within each of the resulting $64$ cells (see Fig.~\ref{fig:DWave:embedding}). By sampling embeddings across the entire QPU, we additionally average over spatial variations in noise, calibration, and hardware performance. Moreover, to mitigate calibration errors and systematic biases, we apply spin-reversal transformations and enable automatic scaling of the local fields and couplings.

\section{Enhancement of Ground-State Probability }
\label{app:DWave:samples_gs}
In section \ref{sec:D-Wave:implementation}, we assessed the performance of the truss-encoded $K_{2,5}$ bipartite MWIS instance on the D-Wave Advantage2\_system2 QPU using the probability of observing the global-minimum basin, defined as the total probability of sampling a computational-basis state within Hamming distance two of the solution state. This metric captures successful tunnelling through the perturbative crossing while remaining robust to subsequent intra-basin excitations.

\begin{figure}[htbp]
    \centering
    \includegraphics[width=\linewidth]{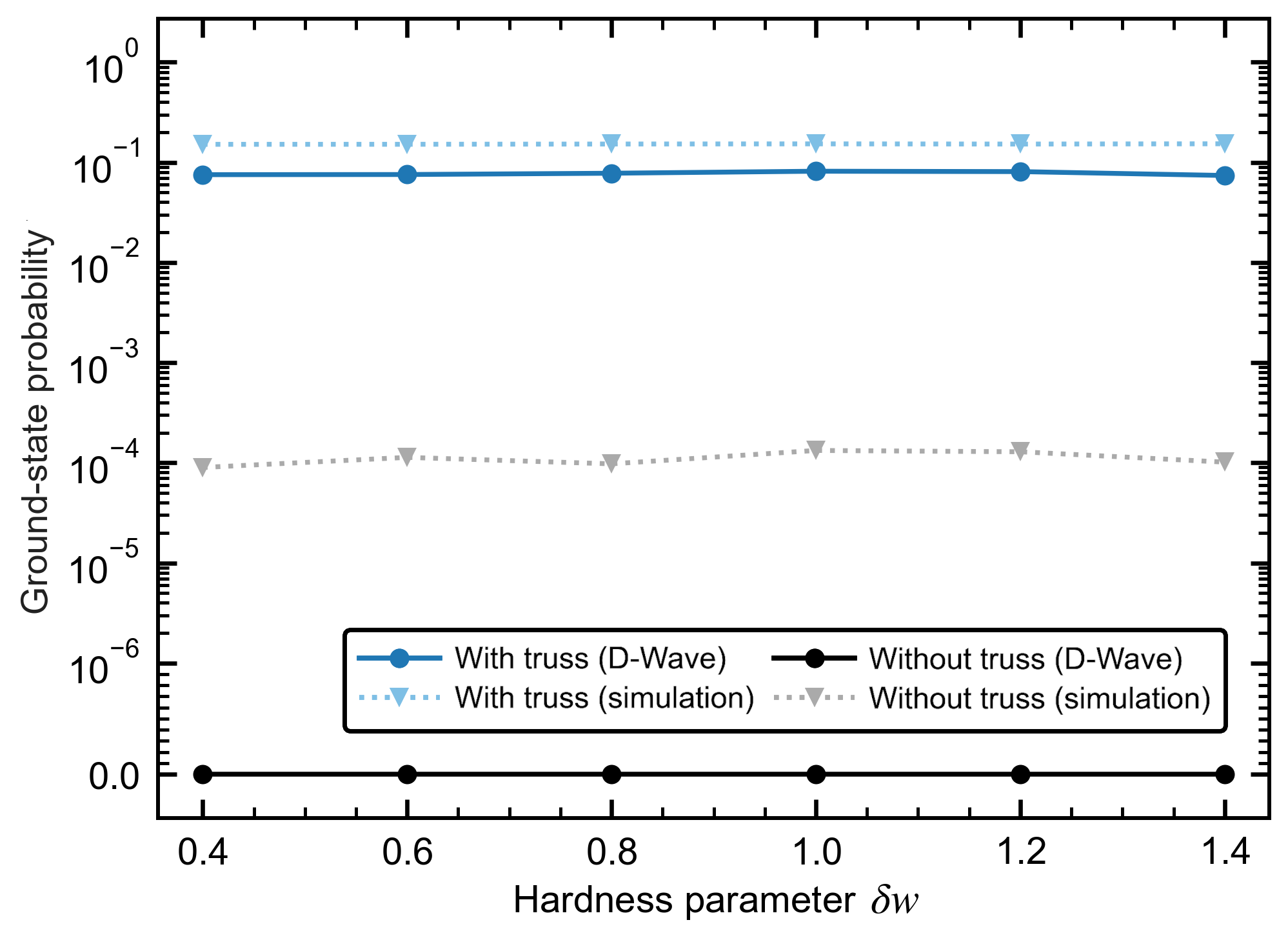}
    \caption{Comparison of the ground-state probability for the $K_{2,5}$ MWIS instance (cf.~Fig.~\ref{fig:DWave:samples}) with ($20$--$27$ physical qubits) and without ($7$ physical qubits) $XX$ coupling  implemented via the truss gadget as a function of the hardness parameter $\delta w$. Solid lines show results obtained on the D-Wave Advantage2\_system2 using $64$ simultaneous minor embeddings and $10{,}000$ QPU samples, yielding $6.4\times10^{5}$ measurements in total with annealing time $0.5\, \mu \mathrm{s}$. Auto-scaling and spin-reversal transformations are used. Dotted lines show ideal state-vector simulations without decoherence.}
    \label{fig:DWave:samples_gs}
\end{figure}

For completeness, Fig.~\ref{fig:DWave:samples_gs} shows the exact ground-state probabilities corresponding to the experiments presented in Fig.~\ref{fig:DWave:samples}(c). Consistent with the probability of observing the system in the ground-state basin, the simulated exact ground-state probability exhibits a substantial enhancement under the truss encoding of approximately three orders of magnitude. This confirms the conclusions drawn in section \ref{sec:D-Wave:implementation}. The exact ground-state probabilities, however, reveal a more pronounced discrepancy between the experimental measurements and the ideal state-vector simulations. For the native instance without the truss gadget, the global-minimum configuration is not observed in any of the experimental samples, whereas the simulations yield an average ground-state probability of approximately $10^{-4}$. In contrast, for the truss-encoded instance, the experimental probabilities are reduced relative to the ideal simulations by only approximately a factor of two. This behaviour is consistent with open-system effects on the D-Wave annealer. In the truss-encoded Hamiltonian, the introduction of auxiliary variables and their associated constraint terms increases the excitation energies associated with single spin-flips in the vicinity of the global minimum. The encoded problem is therefore less susceptible to weak noise-induced transitions out of the ground-state configuration than the native embedding. Consequently, the exact ground-state probability alone does not fully capture whether the system has successfully traversed the perturbative crossing: population may reach the global-minimum basin but subsequently be redistributed among nearby excited states. This observation justifies the use of the problem-dependent ground-state basin probability as the primary success metric in section \ref{sec:D-Wave:implementation}.

\end{document}